\documentclass[twocolumn,showpacs,preprintnumbers,amsmath,amssymb,floatfix]{revtex4}
\usepackage{graphicx}
\usepackage{dcolumn}
\usepackage{bm}
\begin{document}
\def\x{{\mbox{\boldmath$x$}}}
\def\p{{\mbox{\boldmath$p$}}}
\def\q{{\mbox{\boldmath$q$}}}
\def\f{{\mbox{\boldmath$f$}}}
\def\u{{\mbox{\boldmath$u$}}}
\def\v{{\mbox{\boldmath$v$}}}
\def\r{{\mbox{\boldmath$r$}}}
\def\z{{\mbox{\boldmath$z$}}}
\def\y{{\mbox{\boldmath$y$}}}
\def\k{{\mbox{\boldmath$k$}}}
\def\L{{\mbox{\boldmath$L$}}}
\def\unitr{{\mbox{\boldmath$\hat r$}}}
\def\t{\vartheta}
\def\nab{{\bf \nabla}}
\def\eps{{\epsilon}}
\def\ba{\begin{eqnarray}}
\def\ea{\end{eqnarray}}
\def\be{\begin{equation}}
\def\ee{\end{equation}}
\title{Response maxima in modulated turbulence \\
Part II: Numerical simulations}
\author{Anna von der Heydt$^{1,2}$, Siegfried Grossmann$^1$, and 
Detlef Lohse$^2$}
\affiliation{
$^1$Fachbereich Physik, Philipps-Universit\"at Marburg, Renthof 6,
35032  Marburg, Germany\\
$^2$Department of Applied Physics  and J.\ M.\ Burgers Center for
Fluid Dynamics, 
University of Twente, 7500 AE Enschede, 
The Netherlands}
\date{\today}
\begin{abstract}
{Numerical simulations of fully developed turbulence driven by a 
modulated energy input rate or driving force 
are performed within two dynamical cascade models, the GOY 
shell model and a reduced wave vector set approximation of the 
Navier-Stokes equation (REWA). 
The frequency behavior of the system response is 
studied and compared with predictions from a variable range mean-field 
theory, which excludes turbulent fluctuations. In agreement with the 
mean-field approach we find a constant response 
amplitude for low driving frequencies and a $1/\omega$-decay of the 
amplitude for high frequencies. In the mean-field theory, the finite 
cascade time scale had lead to an oscillating behavior of the response 
amplitude as a function of the driving frequency. In the simulations
of both models  we observe the main maximum. 
The higher maxima and minima are completely washed out by fluctuations.}
\end{abstract}
\maketitle

\section{Motivation}
\label{secmotivation}

Many realistic turbulent flows are subject to modulated driving forces, as
e.g. the atmosphere of the earth driven by the periodic heating of the
sun or the pulsed flow through a pipeline. 
Three dimensional
turbulence is characterized by an energy cascade from the outer length
scale, where the forcing acts, to the dissipative scale, where most of
energy is dissipated, see e.g. \cite{pop00,fri95}. 
The down-cascading of energy from large to small
scales takes a characteristic time $\tau$. In a
statistically stationary flow the energy dissipation rate equals the
energy input rate. In a situation with time dependent energy input, on
the other hand, 
this statement will only hold on {\it average}, whereas the
energy dissipation at a certain time $t$ is expected to depend on the
energy input at an {\it earlier} time due to the finite time delay of the
energy transfer. 

In a previous work \cite{hey02} the effect of an energy input rate 
modulated in time, 
\begin{equation}
\label{ein-eq}
e_{in}(t)=\epsilon_0(1+e\sin{\omega t}),
\end{equation}
with a modulation amplitude $e\ll 1$ and a modulation frequency
$\omega$,  
has been studied within a variable range mean-field
theory \cite{eff87}. The response of the system can be observed in the second
order velocity structure function of the flow field at the outer
length scale $L$,
$D_L(t)=\langle\!\langle(\u(\x+\L,t)-\u(\x,t))^2\rangle\!\rangle=6u_{1,rms}$,
which is equivalent to the Reynolds number $Re(t)=u_{1,rms}(t)L/\nu$
of the flow and the total energy 
$E(t)=\langle\!\langle\u^2\rangle\!\rangle/2$ of the system. 
Here, $u_{1,rms}$ is the rms
of one velocity component and $\nu$ is the viscosity.  The response follows the
oscillation of the energy input rate with almost constant modulation 
amplitude at
low frequencies $\omega$ of the energy input rate, whereas  the
response amplitude strongly decreases ($\propto 1/\omega$) at higher
frequencies. The finite energy transfer time $\tau$ plays a
crucial role in this theory. This time $\tau$ is the average time the energy
stays within the system while it is transported by the interaction 
cascade from the large eddies  
towards the small eddies, where it is finally dissipated. This
intrinsic time scale of the system is  a multiple $a$ of order 1 
of the large eddy turnover time $\tau_L$, 
corresponding to the sum over the eddy turnover times on all
scales. $\tau^{-1}$ determines the frequency at 
which the crossover takes place between the regime of 
constant response amplitude and decreasing amplitude. 
In addition, it leads to an oscillating behavior of the system response
with driving frequency $\omega$, where the maxima and minima are at
frequencies connected to the inverse of the energy transfer time
$\tau$. 
In the limit of large frequencies $\omega$, 
the extrema of the response can be
estimated to be at frequencies $\omega_r\simeq n\frac{\pi}{\tau}$,
$n=1,2,3,...$.  

\begin{figure}[b!]
\includegraphics[width=\columnwidth]{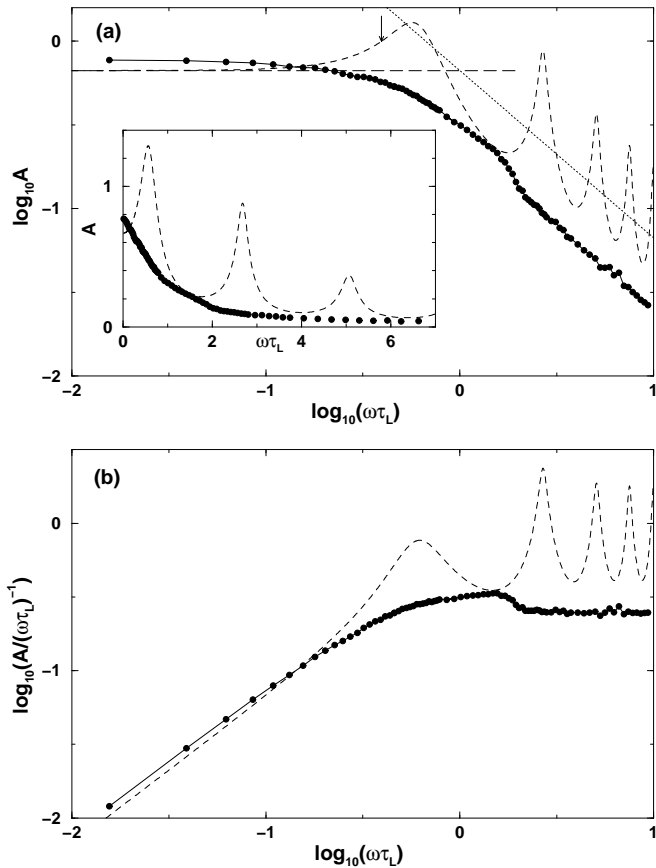}
\caption{
Response amplitude $A$ as a function of
the driving frequency $\omega$ for a modulated energy 
input rate $e_{in}=\epsilon_0(1+e\sin{\omega t})$ calculated within the 
GOY shell model (full dots), see section \ref{ssecgoymode}. 
The modulation amplitude 
is set to $e=0.2$, and the cascade time delay turned out to be 
$\tau/\tau_L=a=2.54$. The stationary Reynolds number is
\mbox{$Re_0=7.1\cdot10^4$}, the viscosity $\nu=1.01875\cdot10^{-4}$, and the 
large eddy turnover time $\tau_L=15.57$. Time and length units are set
by $\nu$, $k_0$ and $F_0$ in GOY.  
Our findings are compared with the response amplitude 
as calculated within the mean-field model with the same $e$ and $\tau$ 
(dashed lines).   
(a) Log-log plot of the amplitude $A$ versus frequency. 
The long-dashed line denotes the low frequency limit of the mean-field 
theory, $A\simeq 2/3$, and the dotted line the high frequency limit, 
$A\propto 2/(3\omega)$.  
The arrow denotes $\omega\tau_L=1/a=0.39$. Near to this 
frequency the crossover takes place in GOY. 
Inset: linear scale plot of the response amplitude.  
(b) Log-log plot of the amplitude compensated by the asymptotic 
amplitude, i.e., $A/(\omega\tau_L)^{-1}$ 
versus frequency. A clear maximum is observed in GOY 
at a frequency near to the maximum of the mean-field amplitude. }
\label{goymode-fig}
\end{figure}
\begin{figure}[b!]
\includegraphics[width=\columnwidth]{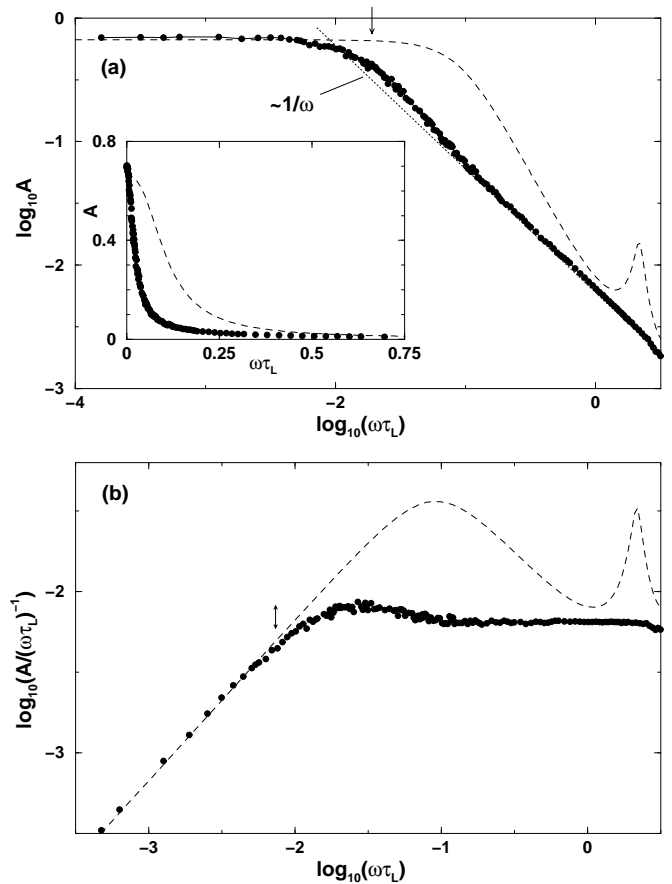}
\caption{
Response amplitude $A$ as a function of
the driving frequency $\omega$ for a modulated energy 
input rate $e_{in}=\epsilon_0(1+e\sin{\omega t})$ calculated within the 
REWA model (full dots), see section \ref{ssecrewamode}. 
The modulation amplitude 
is set to $e=0.3$, and the cascade time scale results to be 
$\tau/\tau_L=a=2.94$. The Kolmogorov constant is found to be $b=83.5$ in 
this simulation instead of $b_{exp}=6-9$. The stationary Reynolds number is
$Re_0=1.2\cdot10^5$, the viscosity $\nu=5\cdot10^{-5}$, and the 
large eddy turnover time $\tau_L=0.063$. Times are measured in units 
of $L_0^{2/3}\epsilon_0^{-1/3}$ in REWA.   
The result is compared with the response amplitude 
as calculated within the mean-field model with the same $e$, $\tau$, and $b$ 
(dashed lines).  
(a) Log-log plot of the amplitude $A$ versus frequency. The dotted 
line is $\propto 1/\omega$. The arrow indicates the mean-field
crossover frequency $\omega_{cross}^{MF}\tau_L=(6/b)^{3/2}=0.019$. 
Inset: linear scale plot of the response amplitude.  
(b) Log-log plot of the compensated amplitude, i.e., 
$A/(\omega\tau_L)^{-1}$ versus frequency. 
A clear maximum is observed in REWA  
at a frequency near to the first maximum of the mean-field
amplitude. The arrow indicates the height of the maximum, i.e., a
deviation from the $1/\omega$-decay by a factor of 1.4 in REWA.}
\label{rewamode-fig}
\end{figure}
\begin{figure}[b!]
\includegraphics[width=\columnwidth]{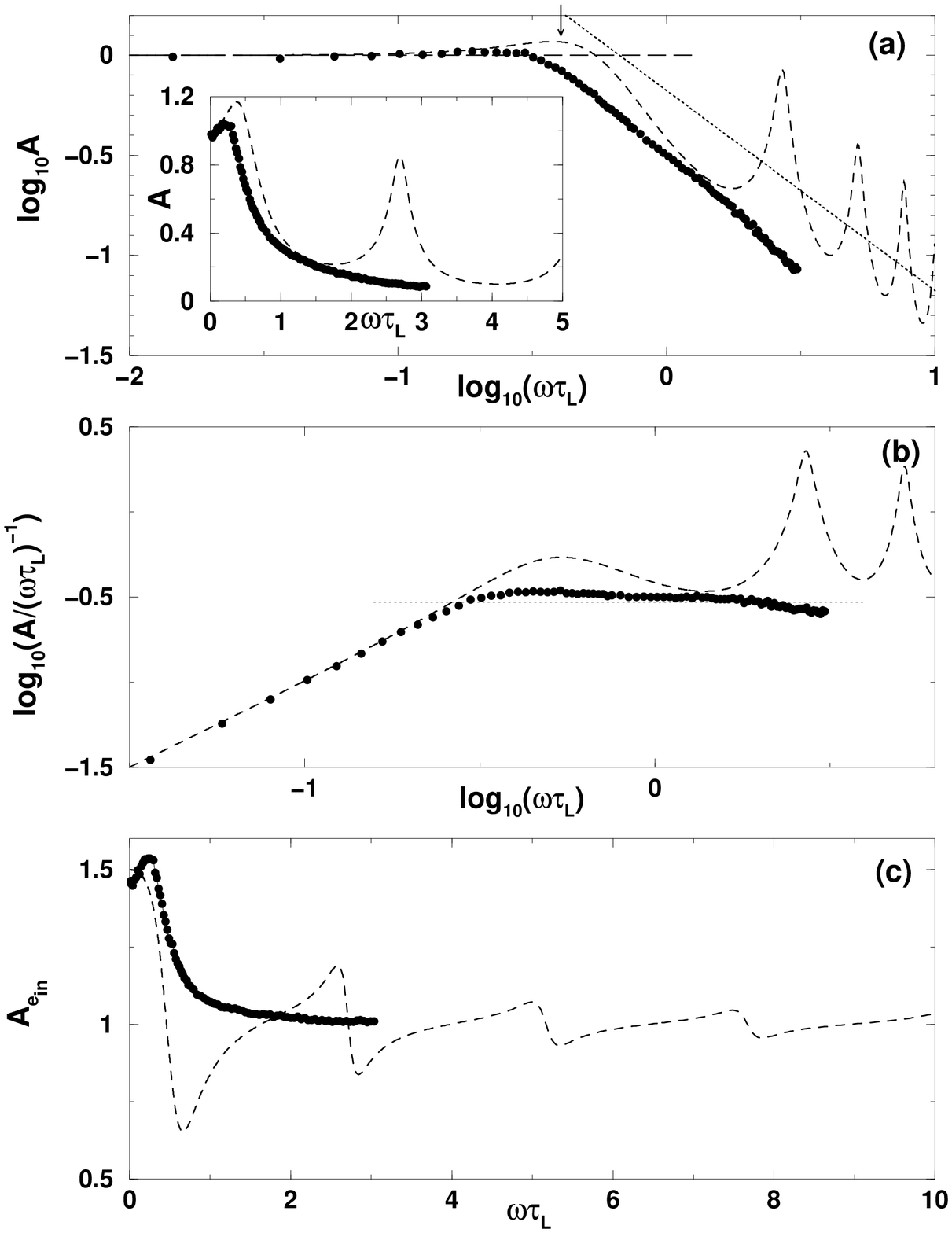}
\caption{
Response amplitude $A$ and amplitude of the energy input rate $A_{e_{in}}$ 
as a function of the driving frequency $\omega$ for a 
modulated driving force  $F=F_{0}(1+e\sin{\omega t})$ calculated 
within the GOY shell model (full dots), see section \ref{ssecgoymodf}.  
The modulation amplitude 
is set to $e_f=0.2$, and the cascade time scale is found to be 
$\tau/\tau_L=a=2.48$. The stationary Reynolds number is
$Re_0=8.6\cdot10^4$, the viscosity $\nu=1.01875\cdot10^{-4}$, and the 
large eddy turnover time $\tau_L=14.5$. 
The result is compared with the response amplitude 
as calculated within the mean-field model with the same $e$ and $\tau$ 
(dashed lines).   
(a) Log-log plot of the amplitude $A$ versus frequency. 
The long-dashed line denotes the low frequency limit of the mean-field theory, 
$A\simeq 1$, and the dotted line the high frequency limit, 
$A\propto 2/(3\omega\tau_L)$. 
The arrow denotes $\omega\tau_L=1/a=0.40$. Near to this 
frequency the crossover takes place in GOY. 
Inset: linear scale plot of the response amplitude.  
(b) Log-log plot of the compensated amplitude, i.e., 
$A/(\omega\tau_L)^{-1}$ versus frequency. The dotted line denotes 
$A/(\omega\tau_L)^{-1}\propto$ const.
(c) Linear scale plot of the energy input amplitude $A_{e_{in}}$ versus 
frequency. The mean-field amplitude as well as the GOY amplitude start
for low $\omega$ with $A_{e_{in}}\simeq1.5$ and merge at
$A_{e_{in}}\simeq1$ for high frequencies. The GOY
amplitude shows only the first main maximum.}
\label{goymodf-fig}
\end{figure}
Recent experiments on modulated turbulence in a cylinder
between two counter rotating disks \cite{cad02} revealed evidence for
the proposed response maxima. In accordance with the predictions from the
mean-field theory \cite{hey02}, for small frequencies 
a constant response amplitude 
 was measured. For large driving frequencies a
$1/\omega$-decay of the velocity response amplitude was observed,
again in agreement with the prediction from our mean-field
approach. Note here, that both the velocity response as well as the
energy response are the same up to a factor of 2, in linear order, cf.
section VI of reference \cite{hey02}. 

In the experiments  the amplitude of the driving force rather than
that of the energy 
input rate is modulated. Since the energy input rate is not 
a controlled quantity any more it can serve to measure the  
response of the system.  Of course, also within the mean-field theory
we can apply a modulated driving {\em force}, see \cite{hey02}. The
main features, the $1/\omega$-decay of the 
energy response amplitude for high frequencies and the constant
response amplitude for low frequencies, pertain. The response maxima are 
only slightly shifted in comparison to the case of a controlled and 
modulated energy 
input rate. In the case of a modulated driving force, as in the
experiments, the energy input rate as a response of the system, also
shows maxima in addition to the mentioned mean features. 
 These are at the same frequencies as the 
maxima of the total energy response amplitude. 

In the mean-field approach, the (intermittent) fluctuations of the
energy and, in particular, of the cascade time $\tau$ are
not present. In experiments and numerical simulations these
fluctuations are of course present, and they 
may lead to broader and less pronounced response maxima and
minima. Therefore, in this paper we shall study the frequency
dependence of the response to a modulated energy input rate into 
a system where turbulent fluctuations are
included. In particular, we shall address the question whether the
response maxima and minima can still be well identified in the
presence 
of fluctuations. 
Furthermore, we not only consider a
modulated energy input rate, but also discuss the slightly different
case of a modulated driving force in order to compare with the above
mentioned experiments.

An appropriate way to numerically study the problem of modulated
turbulence would be a direct numerical
simulation of the Navier-Stokes equation for this specific
time-dependent energy input rate. However, as we need high Reynolds
numbers to achieve fully developed, isotropic, and homogeneous
turbulence and, in addition, need the response  of the system 
as a function of time
for a wide range of driving frequencies, the computational demands 
would be too high. Therefore, we first study the 
problem within a dynamical cascade model of turbulence, 
the Gledzer-Ohkitani-Yamada (GOY) shell
model \cite{gle73,yam87,yam88a,ohk89,kad95,bif03,jen91,bohr}.  
With this model large
Reynolds numbers and enough statistics within a reasonable computing
time for each driving frequency can be achieved. The GOY model has
been successfully used in a study about decaying and kicked turbulence
\cite{hog01}. In addition, to be even closer to a numerical 
Navier-Stokes simulation and to distinguish
between real effects and artifacts of the turbulence model, 
we follow another
approach. We calculate the response of the system to a modulated
energy input rate within a
reduced wave vector set approximation (REWA),
\cite{egg91a,gnlo92b,gnlo94a}, where the Navier-Stokes equation is
solved on a reduced, geometrically scaling subset of wave
vectors. This method is much closer to the Navier-Stokes dynamics than the
GOY-model, as it contains (i) much more modes than GOY, (ii) it solves the
Navier-Stokes equation for those modes and not only a model equation, and
(iii) it is three dimensional.  

Our main results are summarized in Figs. \ref{goymode-fig}, 
\ref{rewamode-fig}, and 
\ref{goymodf-fig}. In  Figs.\ref{goymode-fig} and \ref{rewamode-fig},   
the amplitude $A$ of the energy response is shown as a function of the 
driving frequency for both the GOY model (Fig.\ref{goymode-fig}) 
and the REWA  simulation (Fig.\ref{rewamode-fig}) 
with a modulated energy input rate. This is compared with the results 
of the mean-field model with the corresponding parameters, i.e, the same 
modulation amplitude $e$ and time scale $\tau$. 
In Fig.\ref{goymodf-fig} the results from the GOY model solutions 
are shown for a modulated driving force and compared with 
the mean-field model.  
In all cases we observe a constant amplitude for low 
driving frequencies and a $1/\omega$-decay for high frequencies. 
This can in particular be observed in the compensated plots  
(parts b of all three figures), where $A$, compensated by its
asymptotic amplitude, i.e., $A/(\omega\tau_L)^{-1}$, 
is plotted versus frequency. The $1/\omega$-decay of the response
means, that for fast modulation no response is detectable any
more. The remaining 
dissipation rate is that of the stationary system itself. 

In the mean-field approach, a sequence of response maxima is present
for both types of forcing, starting at a frequency $\omega\propto 1/\tau$.  
In the simulations, this main maximum can also be observed, 
although it is weaker and broader, i.e., it is ``washed out'' 
by fluctuations. The higher order maxima and minima 
are not visible in the simulations, but are completely  
washed out by fluctuations. On the other hand, we emphasize that  
the turbulent fluctuations in the GOY model are strongly overestimated
due to the extreme mode reduction 
in this model. In the REWA simulation, an artificially large Kolmogorov 
constant $b$
is found. Using such large  
$b$ in the mean-field approach also  
leads to a considerable weakening of the first maximum and a 
shrinking of the higher order maxima and minima towards 
very small amplitudes.  

These results will be explained and discussed in detail in this 
paper, which is organized as follows. In the next section we study the
modulated turbulence within the GOY shell model. Before calculating the
response of the system to a modulated energy input rate 
as well as a modulated forcing in
section \ref{ssecgoymode} and \ref{ssecgoymodf} we 
briefly introduce the model and study its stationary 
properties in section  
\ref{ssecgoystat}. In section \ref{secrewa} we present 
our findings on modulated turbulence within
the reduced wave vector set approximation. We summarize our results 
 in section \ref{secconcl}.
 
\section{Modulated turbulence in the GOY shell model}
\label{secgoy}

\subsection{Stationary properties}
\label{ssecgoystat}
The GOY shell model consists of a
set of coupled ODEs for one-dimensional complex velocity modes
$u_n$ \cite{gle73,yam87,yam88a,ohk89,kad95,bif03,bohr,jen91}. 
These modes $u_n$ correspond to velocity {\it differences}
$|\u(\x+\r_n)-\u(\x)|$ on scale $r_n$. $N$ modes are taken
into account, $n=1,2,...,N$, one complex velocity mode per cascade level $n$, defined by the wave numbers $k_n=\lambda^nk_0$ which  are equally spaced on a
logarithmic scale, here, $\lambda=2$. The model equations read:
\begin{eqnarray}
\left(\frac{d}{dt}+\nu
k_n^2\right)u_n&=&i(ak_nu_{n+1}^*u_{n+2}^*+bk_{n-1}u_{n-1}^*u_{n+1}^*
\nonumber\\
\label{goy-eq}
&&+ck_{n-2}u_{n-1}^*u_{n-2}^*)+F\delta_{n,1},
\end{eqnarray}
where $n=1,...,N$, $a=1$, $b=-1/4$, and
$c=-1/2$. These are the traditional parameters. 
We impose boundary conditions on the $u_n$,
i.e., $u_n=0$ for $n<1$ or $n>N$. We use $N=14$ shells, a
viscosity of $\nu=1.01875\cdot 10^{-4}$, and $k_0=2^{-4}$. 
The forcing acts on the
largest scale, i.e., the first shell, $n=1$. 
$F$ is constant, $F=F_0=(1+i)\cdot 10^{-2}$. Together with $\nu$ and 
$k_0$ this sets the time and length units as well as the Reynolds number. 
Equations (\ref{goy-eq}) are integrated using a fourth order
Runge-Kutta scheme with adaptive step size \cite{pre86}.

With the above chosen parameters the GOY dynamics is chaotic \cite{bohr}. 
The system is forced on large scales while most of the energy is dissipated 
on small scales. It reaches a steady state, in which the velocities are 
stochastically fluctuating. In this sense the system has similar properties 
as three dimensional Navier-Stokes turbulence. The scaling behavior of 
structure functions and dissipation has been extensively studied in 
\cite{bif03,jen91,bohr,kad95,ben93,sch95,kad97}. 
The deviations from K41 scaling due to 
intermittency observed in the GOY model are very similar to experimental 
values. Most of the previous studies have been done with 22 or more shells. 
As we use here only 14 shells in order to reduce the computational effort 
we explicitly check some scaling properties of, e.g. the structure functions 
and the energy spectrum in a simulation with constant, stationary forcing 
$F=F_0$ in Equations (\ref{goy-eq}). $N=14$ then turns out to be sufficient.

The Reynolds number of the system can be defined as follows. 
An outer length scale $L$ is given by the 
smallest wave number $k_1$, 
$L=1/k_1$. A typical velocity $U$ is the  
velocity on that scale,  $\langle\!\langle|u_1|^2\rangle\!\rangle_t^{1/2}$. 
The average $\langle\!\langle ... \rangle\!\rangle_t$ is taken over 
time. 
 With these length and velocity scales the Reynolds number of 
the present simulation is 
$Re_0=\frac{UL}{\nu}=8.6\cdot10^4$. 
The simulated time interval is several hundreds of large eddy 
turnover times $|k_1u_1|^{-1}$.  
\begin{figure}[t]
\includegraphics[width=\columnwidth]{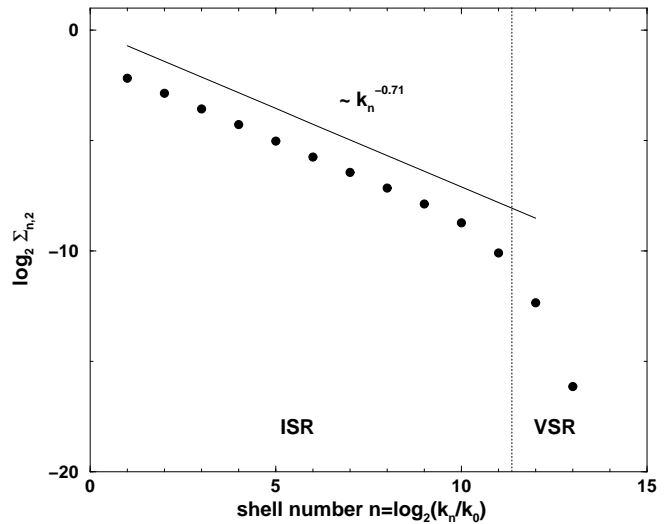}
\caption{Second moment 
$\Sigma_{n,2}=\langle\!\langle|\Im (u_nu_{n+1}u_{n+2}+
\left(\frac{1-\epsilon}{\lambda}\right)u_{n-1}u_{n}u_{n+1})|^{2/3}
\rangle\!\rangle$ (full dots) corresponding to
the second order velocity structure function as a function of
wave number $k_n=2^nk_0$, averaged over a long time interval
($t\simeq1400\tau_L$) and 
with a stationary, constant forcing, $F=F_0$. Between
shell 2 and 9 ISR-scaling behavior, $\Sigma_{n,2}\propto k_n^{-0.71}$ 
(solid line) is
observed, whereas the shells 12-14 form the VSR. $\Sigma_{14,2}$ is
zero by definition. The dotted line 
indicates the middle of the crossover region 
between the ISR and the VSR. }
\label{scalfig}
\end{figure}
\begin{figure}[t]
\includegraphics[width=\columnwidth]{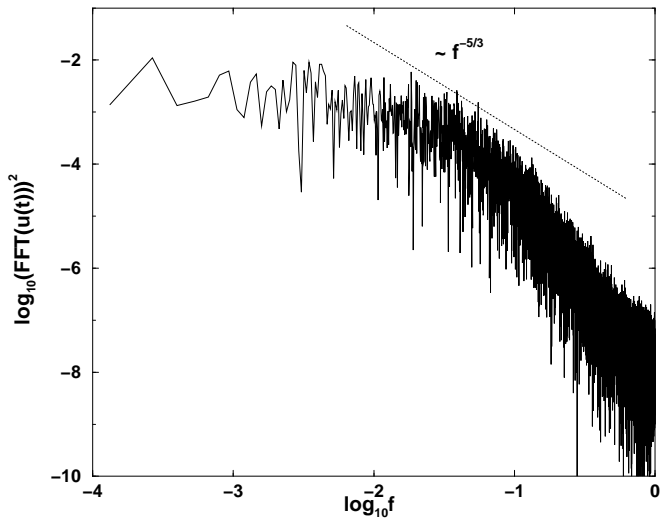}
\caption{Energy frequency spectrum of the
GOY-system with a stationary, constant forcing $F=F_0$. The spectrum
is obtained by fast Fourier transforming (FFT) an actual time series of the
velocity $u(t)=\sum_n\langle\!\langle Re(u_n(t))\rangle\!\rangle$. About one
frequency decade of Kolmogorov scaling $\propto f^{-5/3}$ is
observed. The $k$-space the scaling regime of the spectrum is more
extended, i.e., about three decades, as has been already shown in
Fig.\ref{scalfig}. }
\label{specfig}
\end{figure}
For the second order structure function we use the following method. 
In \cite{kad95} it has been suggested to study the scaling of 
$\Sigma_{n,q}=\langle\!\langle|\Im (u_nu_{n+1}u_{n+2}+
\left(\frac{1-\epsilon}{\lambda}\right)u_{n-1}u_{n}u_{n+1})|^{q/3}
\rangle\!\rangle$ instead of the pure moments of the velocity 
$s_{n,q}=\langle\!\langle|u_n|^q\rangle\!\rangle$ in order to eliminate 
the period 2 and period 3 oscillations which are an artifact of the
GOY model. The second order quantity $\Sigma_{n,2}$,   
corresponding to the second order structure function, is shown in 
Fig.\ref{scalfig}  as a function of wave number index 
$n=\log_2(k_n/k_0)$. 
One clearly observes an inertial subrange (ISR) between
shell 2 and 9, where the second order structure function shows 
scaling with an exponent near to the K41 value 2/3, i.e., $\Sigma_{n,2}
\propto k_n^{-0.71}$ corresponding to $\propto r^{0.71}$ for the
structure function. The scaling exponent is not equal to 
the K41 value, because the model shows intermittency corrections. The
higher wave numbers, i.e., smaller scales, $n=12-14$ belong to the
viscous subrange (VSR), where dissipation takes place. In this range
the viscosity term is dominant and the velocity decays
rapidly with $k_n$, in fact more than exponentially 
\cite{sch95,kad97}. The external
forcing of the flow acts on shell $n=1$, therefore the stirring
subrange (SSR) contains the first shell only. 
The spectrum, which is obtained by fast Fourier 
transforming (FFT) a time series of the
velocity $u(t)=\sum_n\langle\!\langle Re(u_n(t))\rangle\!\rangle$ and
raising it to the power two, can be compared to the energy spectrum
under the assumption of the Taylor hypothesis. This spectrum is shown
in Fig.\ref{specfig} for $N=14$.
We observe about one frequency decade of (nearly) Kolmogorov scaling, 
where the energy decays as $\propto f^{-5/3}$ with frequency $f$. For
$N=14$ the Kolmogorov scaling range is not yet well developed.  For
less shells this region becomes even narrower and the spectrum
shows strong peaks at some intrinsic frequencies. 
When  
reducing the number of shells even more, the velocities relax to a 
stationary value without any fluctuations, i.e., the chaotic behavior 
of the system is lost. 

In conclusion, the GOY model with $N=14$ shells exhibits an inertial 
subrange scaling, although the ISR for the frequency spectrum is only
narrow, but the spectrum in $k$-space has a scaling range of about
three orders of magnitude. This seems acceptable for our goal 
to study the response of the 
system to a modulated driving, because we are only interested in global 
quantities like the total energy $E(t)$ but do not need information on
scale resolved quantities. 

\begin{figure}
\includegraphics[width=\columnwidth]{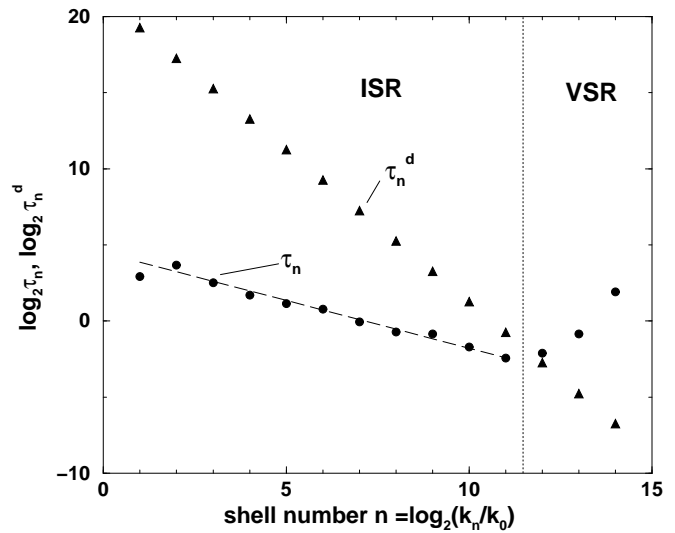}
\caption{Characteristic time scales of the turbulent energy transfer
$\tau_n$ (full dots) and of the viscous dissipation $\tau^d_n$ (full
triangles) as functions of the level number $n$. The dashed line
is a linear fit of $\log_2\tau_n$, $n=2,...,11$, 
and gives $\tau_n^{fit}=22.46\cdot2^{-0.63n}$. 
The shown $\tau_n$ are obtained 
by averaging over about 1400 large eddy turnover times. 
 The dotted line indicates the middle of the crossover region 
between the ISR and the VSR.}
\label{taufig}
\end{figure}

The time scales in the model have been determined as follows. For each
shell $n$ an eddy turnover time $\tau_n$ is defined by \cite{kad95}:
\begin{equation}
\label{taudef-eq}
\tau_n=\frac{1}{|u_n k_n|}.
\end{equation} 
This is also considered as the time scale for the turbulent energy transfer 
through the $n$th level. The time scale
relevant for the energy loss on level $n$ due to  viscosity  is defined as  
\begin{equation}
\label{taudissdef-eq}
\tau^{d}_n=\frac{1}{\nu k_n^2}.
\end{equation}
Both time scales are shown in Fig.\ref{taufig}.  

In the ISR between shell 2 and 9, where the energy transfer  times
are the relevant time scales for the dynamics, the decrease of the
$\tau_n$ with $n$ is near to $\tau_n\propto 2^{-2n/3}$ 
(dashed line in Fig.\ref{taufig}) as
expected for the turnover times of eddies of sizes $r_n/L\propto
({1/2})^n$.  In this range the dissipation time scales $\tau_n^d$
are much larger than the $\tau_n$, meaning that the turbulent energy
transfer is much faster than the viscous dissipation, and therefore the
dominant process. In the VSR instead $\tau_n^d<\tau_n$, i.e., on
average the energy is dissipated by viscosity before it can be
transferred to smaller scales. 

The largest eddy's turnover time is in general defined by the
velocity on the outer length scale $L$, i.e., on the length scale of
the forcing, which in this case is $1/k_1$. However, in this
model, the time $\tau_1$ is disturbed due to finite size 
effects. Therefore, we extrapolate from the turnover times of the 
other shells. A linear fit in Fig.\ref{taufig} for $\log_2\tau_n$
with $n=2,...,11$ leads to $\tau_n^{fit}=22.46\cdot
2^{-0.63n}$. Without intermittency one would have $\tau_n\propto
2^{-2n/3}$; the small deviation corresponds to the intermittent
scaling of $u_n\propto\k_n^{-0.37}$ or an intermittency correction of
$\delta\xi_1=\xi_1-1/3\simeq0.04$.  The extrapolation for
$n=1$ yields for the large eddy turnover time $\tau_L=\tau_1^{fit}=14.5$. 
The time scale corresponding to the energy transfer time $\tau$ used in
the mean-field model \cite{hey02} is the sum over the eddy turnover
times of all energy-input and inertial-range shells, i.e., here
$\tau\simeq\sum_{n=1}^{11}\tau_n\simeq 35.9$. 
The factor $\tau/\tau_L=a$ between the transfer time and the 
large eddy turnover time is then $a=2.48$.

\subsubsection*{Constant energy input rate}
\label{sssec-einconst}
Until now, we have considered a constant forcing $F=F_0$. The resulting 
energy input rate $e_{in}(t)=\langle\!\langle
u_1^*(t)F_0\rangle\!\rangle$ then fluctuates around its mean 
value because of the $u_1^*(t)$-fluctuations.  
In the mean-field theory \cite{eff87} the energy input rate
$e_{in}$ is constant instead. 
For closer comparison we also consider 
another type of forcing in the GOY-model: 
$F(t)=\epsilon_0 u_1(t)/|u_1(t)|^2$. This forcing $F(t)$ fluctuates 
as $u_1(t)$. 
Then the energy input rate is $e_{in}=\langle\!\langle u_1^*
F\rangle\!\rangle=\epsilon_0=const$ by definition. 
The ISR-scaling behavior as
well as the energy spectrum then turn out to be similar 
to the previously discussed
ones with the constant forcing $F=F_0$. The energy transfer time $\tau$
is slightly larger in this case, namely $\tau=39.5$, and 
the large eddy turnover time is $\tau_L=\tau_1^{fit}=15.57$. Again,
the large eddy turnover time is extrapolated from
$\tau_2,...,\tau_{11}$.  
This leads to the factor $\tau/\tau_L=a=2.54$ between the 
total time delay of the energy cascade  and the 
large eddy turnover time. 
In the following sections we will study the time-dependent cases where
either the energy input rate $e_{in}$, i.e., 
$F=\epsilon_0 u_1/|u_1|^2$ (section \ref{ssecgoymode}) or the
 forcing $F=F_0$  is
modulated (section \ref{ssecgoymodf}). 

\subsection{Modulated energy input rate}
\label{ssecgoymode}

In this section we apply a modulated energy input rate to the GOY model, 
i.e., we set the forcing $F=F(t)$ in equations (\ref{goy-eq}) as 
\be
F(t)=\epsilon_0\frac{u_1}{|u_1|^2}(1+e\sin\omega t)
\label{modein-eq}
\ee
with a modulation amplitude $e=0.2$. 
Then, the resulting energy input rate $e_{in}$ is 
\begin{eqnarray}
\label{Eindef-eq}
e_{in}(t)&=&\langle\!\langle u_1^*(t)F(t)\rangle\!\rangle,
\\\nonumber
&=&\epsilon_0(1+e\sin{\omega t}),
\end{eqnarray}
and has a prescribed modulation amplitude $e\epsilon_0$ by definition. 
The total energy 
of the system 
\begin{equation}
\label{Edef-eq}
E(t)=\frac{1}{2}\sum_{n=1}^{14} 
\langle\!\langle u_n^*(t)u_n(t)\rangle\!\rangle,
\end{equation} 
is calculated for a wide range of driving frequencies $\omega$ 
in order to study 
the frequency behavior of the response. 
The brackets 
$\langle\!\langle...\rangle\!\rangle$ denote the ensemble
average. This ensemble average is performed as follows. 
From a long stationary
simulation we collect an ensemble of 1500 starting configurations 
which 
we then let evolve according to
Eqs.(\ref{goy-eq}) but now including the  modulation of the  
forcing $F(t)$, Eq.(\ref{modein-eq}), 
and average over
these 1500 time series. To
ensure that the different realizations can be considered as 
statistically independent, the time delay between the successive
starting configurations for the different realizations is chosen to be 
about  100 large eddy turnover times.  The adaptive step size routine 
controlling the Runge-Kutta integration algorithm for ODEs does not produce 
the same time steps for all time series. To overcome this, we have calculated  
also the same number of equidistant time steps by spline interpolation for all 
time series. 

\begin{figure}
\includegraphics[width=\columnwidth]{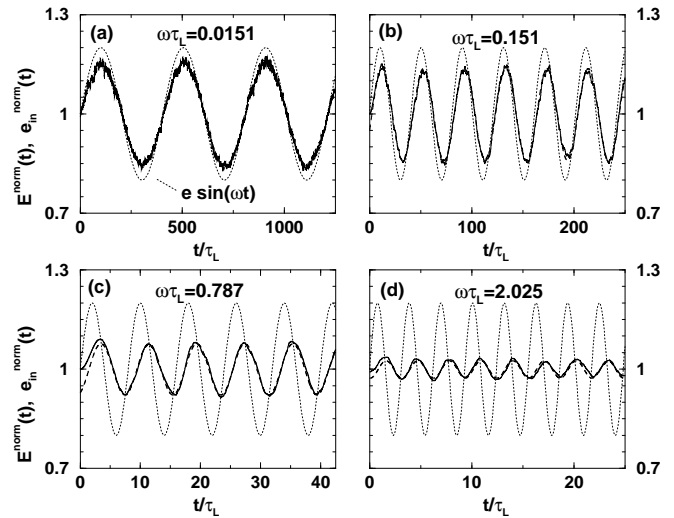}
\caption{Energy input rate $e_{in}^{norm}=e_{in}/\epsilon_0$ (dotted lines)  and energy content 
$E^{norm}$ (solid lines) for four different
modulation frequencies $\omega$ calculated in the GOY model. 
The energy input rate $e_{in}$ is modulated with
a  modulation amplitude of $20\%$  of the constant 
energy input rate $\epsilon_0$, $e=0.2$, according to Eq.(\ref{Eindef-eq}). 
Also included is the fit according to Eq.(\ref{fit-eq}) for the energy 
$E^{norm}$ (dashed lines, indistinguishable 
from the solid lines).  
(a) $\omega\tau_L=0.0151$, 
(b) $\omega\tau_L=0.151$,  
(c) $\omega\tau_L=0.787$, 
(d) $\omega\tau_L=2.025$.}
\label{4ffufig}
\end{figure}
The oscillating response of the system $\Delta(t)$ is then studied in terms of 
the ratio between the energy $E(t)$ with modulated energy input 
 and the energy $E_0(t)$ 
without modulation, namely 
\begin{equation}
\label{goyresp-eq}
E^{norm}(t)=\frac{E(t)}{E_0(t)}=1+\Delta(t).
\end{equation} 
$E$ and $E_0$ both are averaged over 1500 realizations. In spite of
the averaging not only $E$ but also $E_0$ still contains (weak)
fluctuations.  Therefore, we write  $E_0(t)$, as
$E_0$ is still slightly fluctuating around its mean value. $E$ and
$E_0$ contain about the same size of fluctuations. 
Accordingly, the energy input rate $e_{in}$ is normalized by its stationary 
value, $e_{in}^{norm}(t)=e_{in}(t)/\epsilon_0=
1+e\sin{\omega t}$. 

In Fig.\ref{4ffufig} the input rate $e_{in}^{norm}(t)$ and 
the energy $E^{norm}(t)$ 
are plotted for  four different driving frequencies. 
For the two
low frequencies where $\omega\tau_L\ll\tau_L/\tau=1/a\simeq 0.39$, 
the energy follows
the oscillation of the energy input rate with almost constant, but
smaller  amplitude. For higher frequencies the amplitude of the 
deviations of the normalized energy from its stationary value 1 
strongly decrease, and a phase shift with respect to the energy input
becomes visible. The same behavior of the energy has been observed in
the mean-field theory \cite{hey02}.  

To quantitatively access the
frequency behavior of the response amplitude, we calculated time series of 
the total energy $E(t)$ for 85
different driving frequencies varying over almost 3 decades between
$0.012\leq\omega\tau_L\leq 9.3$. The chosen frequencies are  
approximately equally spaced on a logarithmic scale. The 
normalized energy $E^{norm}(t)$ is fitted 
by a function of the form 
\begin{equation}
\label{fit-eq}
E^{norm}(t)=E_{const}+eA\sin{(\omega t+\Phi)},
\end{equation} 
with three free parameters:  $E_{const}$, 
 the amplitude $A$, and the phase shift $\Phi$. 
$E_{const}$ is near to 1 for all frequencies, i.e.,
$E_{const}=1.0022\pm0.0032$. 
The fits (\ref{fit-eq}) are included in 
Fig.\ref{4ffufig} as dashed lines but they are 
mostly indistinguishable from the 
solid lines for the energy itself.

Fig.\ref{goymode-fig} of section \ref{secmotivation} 
shows the amplitude $A$, resulting from the 
fitting procedure, as a function of the dimensionless frequency 
$\omega\tau_L$. $A$ is almost constant for low frequencies and 
has a value of about $2/3$. For higher frequencies the amplitude 
decreases as $\propto 1/\omega$. 
The same features 
have been observed in the mean-field calculations, see dashed lines. 
The long-dashed line in Fig.\ref{goymode-fig} represents the 
low frequency limit of the 
mean-field theory, $A\simeq 2/3$, and the dotted line the high 
frequency limit, $A\propto 2/(3\omega\tau_L)$. 

The crossover between the regime of constant amplitude and the one of 
$1/\omega$-decay of the energy response takes place 
at $\omega_{cross}\tau_L\simeq 1/a\simeq0.39$, which is indicated 
by the 
arrow in Fig.\ref{goymode-fig}. 
In the mean-field approach this crossover is always at $\omega\tau_L=1$, 
independent of the factor $a$ between the large eddy turnover  time and 
the total time scale of the energy transfer. 
In experiments \cite{cad02} the crossover frequency has been 
used to measure the energy cascade time scale. The present simulations 
confirm that this frequency gives the correct order of magnitude for the 
energy transfer time. 

Response maxima, as observed in the mean-field model at frequencies
connected with the inverse energy transfer time, are difficult 
to be identified in
Fig.\ref{goymode-fig}a. There is some structure visible at
$\omega\tau_L\simeq0.31$ and $\omega\tau_L\simeq1.57$. In
Fig.\ref{goymode-fig}b, where the amplitude $A$, compensated by the
asymptotic amplitude $(\omega\tau_L)^{-1}$, i.e.,
$A/(\omega\tau_L)^{-1}$  
is plotted versus frequency, this 
structure becomes more evident, and we see a clear maximum at a frequency 
of about $\omega\tau_L\simeq1.57$. This 
maximum probably corresponds to the mean-field maximum.
Of course, the maximum in GOY is broadened and weakened due to the 
large fluctuations, 
and the higher order maxima and minima are apparently washed out completely. 
\begin{figure}
\includegraphics[width=\columnwidth]{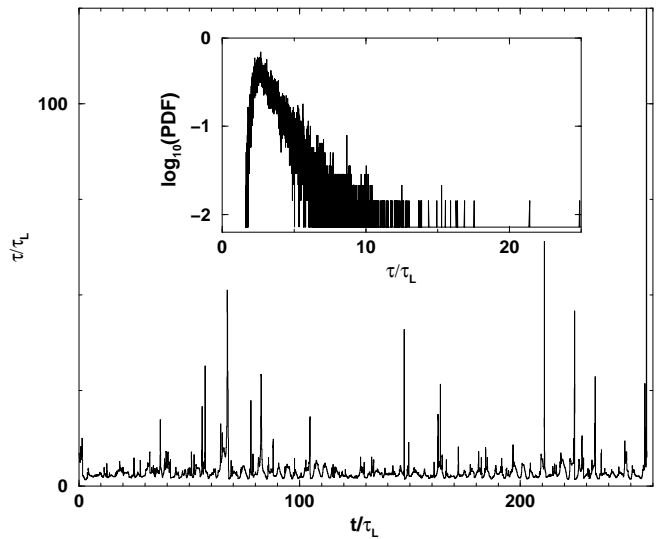}
\caption{Time series of the cascade time $\tau(t)$ for non-modulated 
forcing $F=\epsilon_0 u_1/|u_1|^2$ in the GOY model. 
Strong fluctuations are
observed. Inset: Probability distribution of $\tau/\tau_L$. The mean
is $a=2.54$,  and the width FWHM/2=0.46$a$ giving about 50\% fluctuations.}
\label{tauwidthfufig}
\end{figure}
As in the mean-field theory no fluctuations are included, the energy cascade 
time $\tau$ is considered to be constant. 
However, in the GOY model this assumption is not true, as can clearly be 
seen in Fig.\ref{tauwidthfufig}. Here, a time series of the cascade
time $\tau(t)=\sum_{n=1}^{11}\tau_n(t)$ 
is plotted, computed within the GOY-shell model with
non-modulated forcing $F=\epsilon_0u_1/|u_1|^2$.  The 
inset shows the probability 
distribution of $\tau/\tau_L$. This distribution has its maximum at $\tau/\tau_L=2.39$, almost 
at the mean cascade time $a=2.54$, and a width (FWHM/2) of about 
$0.46 a$. The width is almost half the size of the mean which indicates 
that the transfer time 
fluctuates strongly and therefore we have to expect that 
the response maxima are more or less washed out. 
However, these strong fluctuations are considered as 
an artifact of the GOY-model
and not as a feature of real turbulence. The GOY-model contains only one
velocity mode per cascade level instead of infinitely many modes in
real turbulence. This one-mode approximation leads to an
overestimation of the fluctuation strength. 
In order to confirm this, we performed another simulation with more
modes per level within the reduced wave vector set
approximation (REWA) of the Navier-Stokes equation. This will be
presented in section \ref{secrewa}.

\subsection{Modulated driving force}
\label{ssecgoymodf}

In this section we present further 
results within the GOY model based on a non-fluctuating driving force 
$F_0$ which is regularly  modulated as was $\epsilon_0$ in the
previous section. This case may be more comparable to the experimental
method  in reference \cite{cad02}, because there the 
driving force is modulated. What cannot be modeled with GOY is the
spatial inhomogeneity in the experiments \cite{cad02}. 

In equations (\ref{goy-eq}) we now apply a forcing
\begin{equation}
F=F(t)=F_0(1+e_f\sin\omega t),
\label{modf-eq}
\end{equation}
with a modulation amplitude of $e_f=0.2$. As in section  
\ref{ssecgoymode} we calculate the ensemble averaged 
time series of the energy input rate $e_{in}(t)$, see first line of equation 
(\ref{Eindef-eq}), and the total energy of the system $E(t)$ cf. equation 
(\ref{Edef-eq}), for 89 different frequencies between 
$0.0144\leq\omega\tau_L\leq 3.04$, again logarithmically equally
distributed.  The normalized energy $E^{norm}(t)$ and 
energy input rate $e_{in}^{norm}(t)$  are then 
fitted by a function according to 
equation (\ref{fit-eq}), with the parameters $E_{const}$, $A$, $\Phi$, 
and $e_{in,const}$, $A_{e_{in}}$, $\Phi_{e_{in}}$, respectively.

The amplitudes $A$ and $A_{e_{in}}$ are plotted in 
Fig.\ref{goymodf-fig} of section \ref{secmotivation} as a function of the
dimensionless frequency $\omega\tau_{L}$. Also in this case of a
modulated force, the response amplitude is almost constant for
small frequencies, namely $A\simeq 1$, and 
decreases as $1/\omega$ for high frequencies, see Fig.\ref{goymodf-fig}a 
(full dots). 
Again, the long-dashed line represents the low 
frequency limit of the mean-field theory for $A$ (which is 
$A\simeq 1$ in this case) as well as the dotted line the high 
frequency limit. As in the previous section, the crossover frequency 
between the two regimes is determined by the energy transfer 
time, i.e.,  
$\omega_{cross}\tau_L\simeq 1/a\simeq0.40$ (as $a=2.48$ in this case), 
which is marked by the small arrow in 
Fig.\ref{goymodf-fig}a. 
 The amplitude of the energy input rate $A_{e_{in}}$ starts
with a value of about $1.5$ for low frequencies and merges towards 
$1$ for high frequencies, see Fig.\ref{goymodf-fig}c. This indicates
that at very large frequencies the velocity is not oscillating any
more as it only feels a mean constant force. The oscillations of the
energy input rate are then only a consequence of the oscillation of
the driving force $F$. 
 In the mean-field theory we have observed the same trend 
for both amplitudes. The corresponding  mean-field results are 
included as dashed 
lines. 

Both amplitudes $A$ and $A_{e_{in}}$  show a maximum at a
frequency near to the crossover frequency 
$\omega\tau_L\simeq\tau_L/\tau=a^{-1}=0.40$. 
In the compensated plot Fig.\ref{goymodf-fig}b, where
$A/(\omega\tau_L)^{-1}$  is 
plotted as a function of frequency, a clear deviation from the  
dotted line representing $A\propto1/\omega$ can be observed. 
At this frequency, the mean-field theory  
predicts a first maximum
for the energy response amplitude (Fig.\ref{goymodf-fig}a) and 
a maximum directly followed by 
a minimum for the energy input rate, see Fig.\ref{goymodf-fig}c.  
This frequency is 
connected  to the energy transfer time $\tau$. 
In the mean-field model further maxima of the energy response and wiggles 
of the energy input rate are observed at multiples of this frequency. 
However, in the GOY model all further maxima and minima are washed out 
presumably because of the strong fluctuations. 
The fluctuations of the energy cascade time scale are found to be
similar as in the case of a fluctuating force 
shown in Fig.\ref{tauwidthfufig} in the previous section. 


\section{Modulated turbulence in a reduced wave vector set approximation}
\label{secrewa}

As was pointed out in the introduction of this chapter a full numerical 
simulation   of the Navier-Stokes equation for modulated turbulence at high 
Reynolds numbers is still not possible or requires low Reynolds
numbers.  Therefore, we first have considered 
 the GOY shell model. This model correctly describes many features of 
turbulence, however, due to the one-mode approximation in each cascade level, 
it contains various artifacts. 
Namely, it strongly overestimates the strength of the fluctuations. 
The aim of this section is to study the problem of modulated turbulence within 
another model, the reduced wave vector set approximation (REWA)  
\cite{egg91a,gnlo92b,gnlo94a}, 
 which 
is much closer to the Navier-Stokes equation than the GOY model and
contains much more modes per cascade level. Of course, as compared to
full numerical simulations  
of the Navier-Stokes equation, it still contains a mode reduction in
order to make   
the computational effort reasonable for the desired high Reynolds numbers. The 
present approximation has been introduced and extensively studied in 
\cite{egg91a,gnlo92b,gnlo94a}
. Here, we use it together with a time-dependent driving. For
completeness we briefly explain the approximation before we present the 
results with modulated driving.

\subsection{The reduced wave vector set approximation}
\label{ssecrewadescribe}  

The velocity field $\u(\x,t)$ is Fourier transformed into plane waves,
$\u(\x,t)=\sum_{\p}\u(\p,t)e^{i\p\cdot\x}$. Periodic boundary
conditions are applied on a periodicity volume $(2\pi L_0)^3$. The wave
vectors $\p$ are given by $\p=(p_i)=(n_iL_0^{-1})$, with
$n_i=0,\pm1,\pm2,...$. In order to efficiently deal 
with the large number of modes
involved, the reduced wave vector set approximation 
selects a limited number of modes by admitting only a geometrically
scaling  subset
$K=\bigcup_l K_l$ of wave vectors, i.e., $\u(\x,t)=\sum_{\p\in
K}\u(\p,t)e^{i\p\cdot\x}$. On this subset $K=\{\p_n^{(l)}, n=1,...,N,\  
l=0,...,l_{max}\}$  the Navier-Stokes
equation for incompressible flow,  
\ba
\label{rewa-eq}
\frac{d}{dt}u_i(\p_n^{(l)})&=&-\nu (\p_n^{(l)})^2u_i(\p_n^{(l)})+f_i(\p_n^{(l)})\\
-i&M_{ijk}&(\p_n^{(l)})\sum_{\q_1,\q_2\in K,\q_1+\q_2=\p_n^{(l)}}
u_j(\q_1)u_k(\q_2),
\nonumber
\ea
together with the continuity equation, 
\be
\p_n^{(l)}\cdot\u(\p_n^{(l)})=0, 
\ee
is solved.
$M_{ijk}$ is the coupling matrix, $M_{ijk}(\p)={1\over
2}(p_jP_{ik}^\perp(\p)+p_kP_{ij}^\perp(\p))$, where $P_{ij}^\perp(\p)$
is the orthogonal projector to $\p$. 
The subset $K$ consists of a basic subset
$K_0=\{\p_n^{(0)}, n=1,...,N\}$ together with its scaled replicas
$\p_n^{(l)}=2^l\p_n^{(0)}$, $l=1,...,l_{max}$. 
In the present simulation we take $N=74$
wave vectors $\p_n^{(0)}$,
\ba
K_0=&&\{\pm(2,2,2),\pm(-1,2,2),\pm(-2,1,1),\pm(3,0,0),
\nonumber\\
&&\pm(4,1,1),\pm(4,-2,1),\pm(-3,3,3),\pm(-5,1,1),
\nonumber\\
&&\pm(4,4,1),\pm(3,3,0),\pm(1,1,10),\pm(-10,5,5)\nonumber\\
&&+\mbox{permutations}\}
\ea
These wave vectors $\bigcup_lK_l$ 
are chosen such that they span a wide range of
length scales, but still dynamically interact to a good degree. For
the Navier-Stokes equation this means that as many Navier-Stokes
interactions $\p=\q_1+\q_2$ as possible between the wave vectors in
$K$ are allowed for. The
largest eddies of the order $L_0$ are represented by the wave vectors in
$K_0$, whereas the subsets $K_l$ contain wave vectors of smaller and
smaller eddies. The choice of the smallest eddies, i.e., the value of
$l_{max}$, depends on the kinematic viscosity $\nu$. $l_{max}$ and
$\nu$ are adjusted such that the velocity amplitudes
$\u(\p_n^{(l_{max})},t)$ of the smallest eddies are almost zero. In this
simulation, $\nu$ is chosen as $\nu=5\cdot 10^{-5}$ and the number of 
levels as $l_{max}+1=9$. 

To maintain the turbulent flow we apply a forcing as in 
\cite{egg91a,gnlo92b,gnlo94a}:
\ba\label{rewaforce-eq}
\f(\p,t)\!=\!
\left\{ \begin{array}{cc}
\epsilon_0\frac{\u(\p,t)}{\sum_{\q\in K_{in}}|\u(\q,t)|^2}(1+e\sin\omega t), 
& \p\in K_{in}\\[1.5ex]
0, & \p\not\in K_{in}.
\end{array} \right.
\ea
The subset $K_{in}$ of $K_0$ by choice 
contains the wave vectors with the
three smallest lengths. $K_{in}$ contains 14 vectors, namely
$K_{in}=\{\pm(2,2,2),\pm(-1,2,2),\pm(-2,1,1),\mbox{+permutations}\}$. 
In reference \cite{gnlo92b} it has been shown that the statistics of
the solutions of 
the equations of motion do not depend on the
particular choice of $K_0$. 
This forcing corresponds to the same type of forcing, which has
already 
been applied to the GOY model in section \ref{ssecgoymode}. It enforces
the energy input rate to be modulated: 
\be
\label{einrewa-eq}
e_{in}=\langle\!\langle\sum_{\p_n^{(l)}\in
K}\u^\ast(\p_n^{(l)})\cdot\f(\p_n^{(l)})\rangle\!\rangle
=\epsilon_0(1+e\sin\omega
t). 
\ee 

Eq.(\ref{rewa-eq}) is a set of $3N(l_{max}+1)$ coupled ODEs for the
complex mode amplitudes $u_i(\p_n^{(l)})$ which is
numerically solved within the Burlisch-Stoer integration scheme with
adaptive step size \cite{pre86}. Length scales are
measured in units of $L_0$ and time scales in units of 
$L_0^{2/3}\epsilon_0^{-1/3}$. A Reynolds number can be 
defined as follows: The wave length $\lambda$ of the smallest wave vector 
gives an external length scale $L=2\pi/\sqrt{6}$,  and a typical velocity 
on that scale is determined  by the rms of one 
velocity component, $u_{1,rms}$. Then, in our case, the Reynolds number is 
$Re=\frac{u_{1,rms}L}{\nu}=1.234\cdot 10^5$ because from the simulations we
obtain $u_{1,rms}=2.405$.  

The main features of fully developed turbulence as irregular velocity
signals, characteristic scaling of structure functions, etc. are well described
within this approximation, as has been shown in
\cite{gnlo92b,gnlo93c,egg91a}. The REWA solutions show small scale
intermittency, which is produced by the competition between down-scale
energy transport and viscous dissipation on the small scales
\cite{gnlo92b,gnlo93c}. Other mechanisms leading to intermittency in
turbulence as e.g. nonlocal interactions between wave vectors 
 are underestimated in
this approximation \cite{gro96}. 
The down-scale energy transport in the REWA fluid is less effective than in 
real turbulence, because in this approximation the larger wave vectors are 
more and more thinned out \cite{gnlo94b}. 
This is in contrast to the case of the complete set of wave vectors
(e.g. in full grid simulations) where the density of states increases
$\propto p^2$,  whereas in the reduced wave 
vector set $K$ the number of admitted wave vectors decreases as $1/p$  
\cite{gnlo94a}.  In reference \cite{gnlo94b} it has been shown that this reduced energy 
transport leads to an overestimation 
of the Taylor Reynolds number of the system as well as the 
Kolmogorov constant $b$, defined by $D(r)=b(\epsilon r)^{2/3}$, 
by roughly one order of magnitude for our choice of $N$. In the 
present simulation we obtain $b=83.5$ instead of
$b=6-9$ as in experiments \cite{my75,sre95,sre97}. Since $D(L)$ is the
energy density $\propto\langle\!\langle\u^2\rangle\!\rangle$ of the
fluctuations in the fluid system, the large $b$ value indicates that
even in the REWA approximation the strength of the fluctuations is
highly overestimated. The 
large Kolmogorov constant  
will change the relevant time scales in the system, 
as will be shown in section \ref{ssecrewamode}. 

The characteristic time scale for the turbulent energy transfer on
scale $l$ can be estimated as:
\be
\label{taurewadef-eq}
\tau(l)=\frac{1}{p^{(l)}u^{(l)}_{rms}},
\ee
where $p^{(l)}$ denotes the mean wave number on scale $l$, i.e., it is
the
mean inverse eddy size in $K_l$. As in the GOY model, 
Eq.(\ref{taudissdef-eq}), the time 
scale of viscous dissipation is $ \tau_d=[\nu (p^{(l)})^2]^{-1}$. Again, 
in the ISR $\tau(l)>\tau_d$, whereas in the VSR $\tau_d>\tau(l)$. 
From a simulation with stationary forcing, i.e., $e=0$ in Eq. 
(\ref{rewaforce-eq}), 
the time delay of the energy down-transport $\tau_{sum}$ 
is then estimated by the sum of 
all $\tau(l)$ in the ISR, $\tau_{sum}=\sum_{l\in
ISR}\tau(l)\simeq0.186$. 
The largest of these $\tau(l)$, on the largest scale, $\tau(0)=0.0632$ can 
be regarded as a large eddy turnover time $\tau_L$. Thus, 
$\tau_{sum}=2.94\tau_L$, and the factor between the cascade time scale
and the large eddy turnover time is $\tau_{sum}/\tau_L=a=2.94$.  

\begin{figure}
\includegraphics[width=\columnwidth]{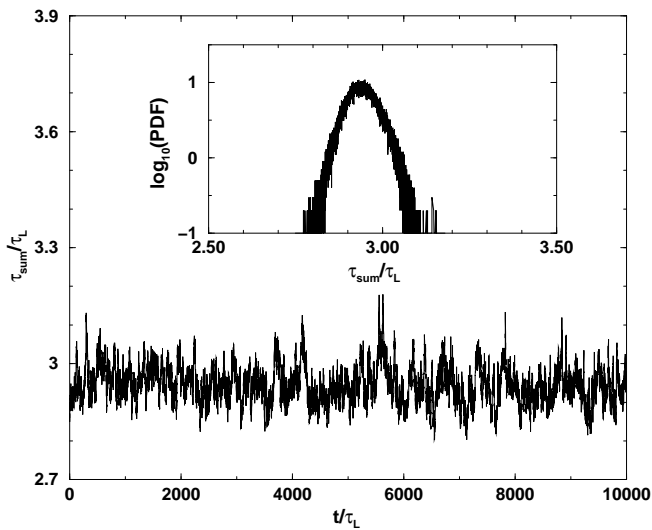}
\caption{Time series of the cascade time $\tau_{sum}(t)$ in REWA. 
Inset: Probability distribution of $\tau_{sum}/\tau_L$. 
The fluctuations are 
considerably smaller than in the GOY model. Note the different scales in this 
figure and Fig.\ref{tauwidthfufig}. 
The mean is $\tau_{sum}/\tau_L=a=2.94$, and the width $FWHM/2=0.02a$ 
giving about 2\% fluctuations.}
\label{tauwidthrewafig}
\end{figure}
As we have seen in the GOY model, Fig.\ref{tauwidthfufig}, the energy
transfer  time
 is strongly fluctuating. We attributed these strong fluctuations
to the one-mode per level approximation of the GOY
model. Fig.\ref{tauwidthrewafig} shows a time series of the energy
transfer time $\tau_{sum}(t)$ (use Eq.(\ref{taurewadef-eq}) with
$u^{(l)}(t)$) in the present reduced wave vector set
approximation (REWA), and, in the inset, the distribution of this time
scale. Clearly, the fluctuations are much weaker than in the GOY
model; they are zero
in the mean-field approximation. 
The distribution is centered around {\bf $\tau_{sum}/\tau_L=a$} with
a width (FWHM/2) of about {\bf $0.02a$}.

\subsection{Modulated energy input rate}
\label{ssecrewamode}

\begin{figure}
\includegraphics[width=\columnwidth]{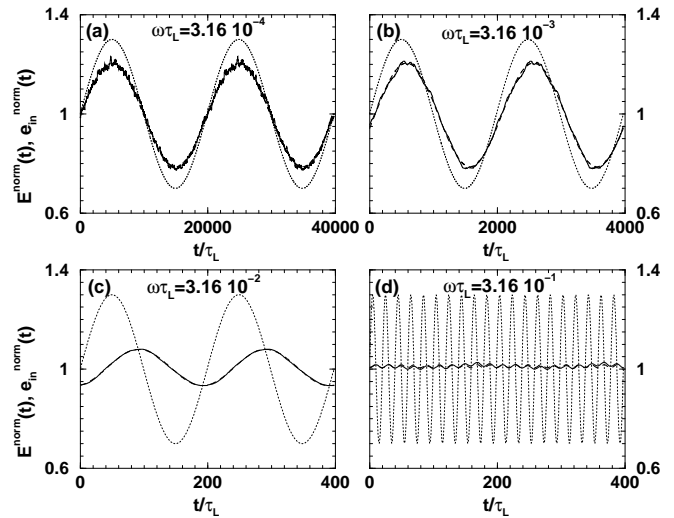}
\caption{Energy input rate $e_{in}^{norm}=1+e\sin{\omega t}$ 
(dotted lines)  and energy
$E^{norm}$ (solid lines) for four different
modulation frequencies $\omega$ as calculated in the REWA simulation. 
The energy input rate is modulated with
a  modulation amplitude of $30\%$  of the constant 
energy input rate, i.e., $e=0.3$ in Eq.(\ref{rewaforce-eq}). Also
included is the fit to the energy data cf. Eq.(\ref{fit-eq}) as dashed
lines but these are in all cases indistinguishable from the solid lines. 
The averaged 
time series of $E^{norm}$ are repeated once for better visibility.
(a) $\omega\tau_L=3.16\cdot10^{-4}$, 
(b) $\omega\tau_L=3.16\cdot10^{-3}$,  
(c) $\omega\tau_L=3.16\cdot10^{-2}$, 
(d) $\omega\tau_L=0.316$. For larger $\omega\tau_L$ the energy is
indistinguishable from 1 on this scale. The crossover to the
$1/\omega$-decay regime is in this simulation at
$\omega\tau_L\simeq 1.1\cdot10^{-2}$, between the frequencies of b)
and c).}
\label{4frewafig}
\end{figure}

The response of the system to a modulated driving force,
cf. Eq.(\ref{rewaforce-eq}), is calculated now  
in terms of the total energy
of the system
\be
\label{Edefrewa-eq}
E(t)=\frac{1}{2}\langle\!\langle\sum_{l=0}^{l=l_{max}}\sum_{\p\in
K_l}|\u(\p,t)|^2\rangle\!\rangle.
\ee
The modulation amplitude of the energy input rate $e_{in}$ 
(Eq.(\ref{einrewa-eq})) is chosen
as $e=0.3$. 
The average $\langle\!\langle...\rangle\!\rangle$ is performed as follows. 
We average over 25 to 30 realizations, which are obtained from
Eqs.(\ref{rewa-eq}) with different starting values.  
The delay between the different starting values is one driving period. 
For the higher frequencies the period $2\pi/\omega$ 
becomes too small to ensure statistical independence of the different 
realizations. Then, we chose the delay between the successive starting
values of the different realizations 
to be several driving periods  
such that it is at least $60 \tau_{sum}$. 
The response is calculated for 150 (approximately equally spaced on a
logarithmic scale) 
frequencies between {\bf $0.00016\leq\omega\tau_L\leq 3.0$}. 
The energy is normalized by $E_0$, calculated from a
stationarily forced solution with $e=0$ and averaged as $E$. The
oscillating response 
$\Delta(t)$  of the
system is defined in the same way as for the GOY model, see
Eq.(\ref{goyresp-eq}). 
Then, as for the GOY calculations, the averaged and normalized signals $E^{norm}(t)$
are fitted with Eq.(\ref{fit-eq}). The fit parameter $E_{const}$ is again near 
to 1 for all frequencies, $E_{const}=1.0064\pm0.0065$. 
In Fig.\ref{4frewafig} the time averaged
responses and the normalized energy input rates are plotted for four
different driving frequencies. Also the fits according to
Eq.(\ref{fit-eq}) have been included as dashed lines but  are
indistinguishable from the solid lines for the energy signal itself. 

We observe in Fig.\ref{4frewafig} for REWA 
the same features as in Fig.\ref{4ffufig} for the GOY
model and in Fig.1 of reference \cite{hey02} for the mean-field
model. For the two lower frequencies the response amplitude remains
almost constant and is about $2/3$ of the amplitude of the energy
input rate,  
whereas for the two higher frequencies the response amplitude strongly
decreases. This trend becomes more clear in
Fig.\ref{rewamode-fig}a in section \ref{secmotivation},  
where the amplitude of the response --
determined from the fit (\ref{fit-eq}) -- is shown as a function of the driving
frequency (full dots). 
For low driving frequencies $A\simeq 2/3$, whereas for high 
frequencies the amplitude decreases as
$1/\omega$. 
The crossover between the regime of constant amplitude and the one of 
$1/\omega$-decay takes place at $\omega_{cross}\tau_L\simeq0.011$, 
i.e., at a much smaller frequency than expected from the original case 
of the mean-field theory in which the crossover was at 
$\omega_{cross}^{MF}\tau_L=1$ with $b=6$. We understand this as follows. 
In section \ref{ssecrewadescribe} 
it was mentioned that the Kolmogorov constant in the REWA simulation 
is $b=83.5$ instead of $b=6-9$ as in experiments. In the figures of the 
 mean-field approach 
\cite{hey02} we have set $b=6$. 
The mean-field solution for a general $b$ revealed that the
crossover frequency decreases with increasing $b$ while the positions
of the response
maxima are left unchanged. 
For $b=83.5$ the mean-field crossover 
frequency is at $\omega_{cross}^{MF}\tau_L=(6/b)^{3/2}\simeq 0.019$ in
close agreement  to what we observe in the REWA simulations. The response 
amplitude calculated from the mean field model with $b=83.5$ is included 
in Fig.\ref{rewamode-fig}a as dashed line. 
Apart from the changed crossover frequency we 
observe that, in the mean-field calculations,  
the first response maximum at $\omega\tau_L\simeq 0.1$ is 
considerably smaller and broadened  as compared to the case with $b=6$,
given as the dashed line in Fig.\ref{goymode-fig}a. In agreement with
this, our  REWA simulations (with a value $b=83.5$) 
show a broad maximum  in the response amplitude at 
$\omega\tau_L\simeq 0.028$. This means it occurs 
at a similar frequency as the mean 
field model. The maximum becomes more clear in the compensated plot, 
Fig.\ref{rewamode-fig}b, where $A/(\omega\tau_L)^{-1}$ is 
shown as a function of 
frequency. 
There, we observe  a deviation from the $1/\omega$-decay of 
the amplitude by a factor 1.4 at the maximum in the REWA 
simulations. The mean field maximum has a height of 2.8.
The subsequent maxima and minima in the mean field model 
occur at frequencies where the amplitude is already very small 
($A\leq 10^{-2}$) because the crossover to the $1/\omega$-regime 
takes place at a much smaller frequency whereas the response maxima
stay at the same frequencies as for a smaller $b$.
Therefore, the higher order maxima are not visible in the REWA simulations. 
The cascade time shows about  2\% fluctuations in REWA as shown in Fig.
\ref{tauwidthrewafig}. However, at small response amplitudes these 
fluctuations are already 
large enough to wash out the higher order response maxima. 

In conclusion, the REWA system reproduces qualitatively the features 
of modulated turbulence as predicted by the mean-field model including 
the first response maximum. The latter is considerably weakened due to 
the large Kolmogorov constant in REWA.  Another  consequence of the large 
$b$ is, that the crossover between constant amplitude and $1/\omega$-decay 
is shifted towards smaller frequencies, and therefore the higher order 
maxima and minima are already at very small amplitudes where the
fluctuations in the cascade time scale  are finally large enough to
wash them out. We cannot clarify, at present, how close the response
in direct numerical simulations -- which lead to an order of magnitude
smaller $b$ and thus have much smaller fluctuations -- will come to the
mean-field features, but we expect a clearly visible first maximum at
least. 


\section{Conclusions}
\label{secconcl}

We have simulated the response of modulated turbulence within two numerical 
models. Namely, we have used the GOY shell model and the reduced wave vector 
set approximation of the Navier-Stokes equation (REWA). 
The results are compared 
with predictions from a mean-field theory. For a modulated energy input rate 
this mean-field theory had predicted  a constant response amplitude for low 
frequencies and a $1/\omega$-decay for high driving frequencies. In addition, 
at certain frequencies connected with the energy cascade time scale, a
sequence of maxima 
and minima of the response amplitude is observed. 

Both numerical models well reproduce the basic trend, i.e., the constant 
amplitude for small $\omega$ and the $1/\omega$-decrease for large $\omega$. 
The main response maximum can be observed in both numerical models, although 
it is weakened due to fluctuations. The higher order maxima and minima as 
predicted by the mean-field theory cannot be identified in the simulations. 
Obviously, they are strongly washed out by fluctuations.  
In the GOY model the large  fluctuations are explicitly visible in a
broad cascade time distribution. We attribute these fluctuations to
the strong mode reduction in  the model, i.e., they are an artifact of
the model properties and not a 
feature of real turbulence. In  the reduced wave vector set approximation 
of the Navier-Stokes equation these fluctuations are much weaker, and, 
we believe, more realistic for real turbulence, as more modes are taken 
into account. 
However, due to 
an overestimated Kolmogorov constant $b$  
in REWA the higher order maxima are considerably reduced,  
and therefore washed out by the fluctuations although being smaller
than the GOY-fluctuations. Therefore, we 
believe, that 
in real turbulence with a realistic Kolmogorov constant  
 the first maximum should be clearly 
observable and possibly also the higher order extrema in  
the response. Thus, the predictions of the mean-field 
model, which excludes {\it all} fluctuations, might be quite
reasonable for real turbulence. 
To further study the response maxima numerically, it is necessary 
to perform full 
numerical simulations of the Navier-Stokes equation, as then all relevant 
time scales including their fluctuations are reproduced realistically,
which turned out to be essential for the 
observation of the response maxima.   

Recent experiments on modulated turbulence \cite{cad02} revealed evidence 
for the response maxima. These experiments may be more comparable to
the case of a modulated {\em force} instead of a modulated energy input rate. We have
studied this case also within the mean-field model and have found
basically the same behavior of the energy response as for a modulated
energy input rate. In addition, the amplitude of the energy input rate
showed ``wiggles'' at the same frequencies where the energy response
had maxima. In the experiments the response maxima were measured in
the energy input rate, which can be regarded as a response of the
system as well in this case. Also the constant amplitude for low
driving frequencies and the $1/\omega$-decay of the velocity response
-- which in leading order is corresponding to a $1/\omega$-decay of
the energy response, as well -- have been observed in the experiments. 
Here, we have studied the case of a modulated driving 
force within the GOY shell model. Also in this simulation, 
the response amplitude behaves 
basically as in the mean-field model, i.e., it decreases as $1/\omega$. 
The energy response amplitude as well as the amplitude of the energy
input rate show the main maximum. Due to the above mentioned 
large fluctuations all higher order maxima are washed out in the GOY
model.  

There are two regimes in the frequency behavior of the response amplitude, 
namely a constant amplitude at low $\omega$ and a decreasing amplitude at 
high $\omega$. The present simulations give further confidence that the 
crossover frequency between these two regimes gives the correct order of 
magnitude of the cascade time scale, i.e., in experiments it can be used 
to measure this time scale as suggested in reference \cite{cad02}. 

Both models in the present study were able to reproduce the main features of 
the frequency behavior of the response amplitude in modulated turbulence 
as predicted by the mean-field model, however, both also have their 
shortcomings, which prevent us from correctly predicting the behavior 
of real turbulence in all quantitative details.   
Therefore, we believe, that it is worth to further study modulated 
turbulence numerically as well as experimentally. 

\noindent
{\bf Acknowledgments:} 
The work is part of the research  program of the Stichting voor 
Fundamenteel Onderzoek der Materie (FOM), which is financially supported 
by the Nederlandse  Organisatie voor Wetenschappelijk Onderzoek (NWO).
This research was also supported 
by  the German-Israeli Foundation (GIF) 
and by the European Union under contract HPRN-CT-2000-00162.

\vspace{-0.5cm}

\end{document}